\documentclass[a4paper]{jpconf}
\usepackage{graphicx}
\usepackage{citesort}
\usepackage{bm}

\usepackage[usenames]{color}

\newcommand\be{\begin{equation}}
\newcommand\ba{\begin{eqnarray}}
\newcommand\ee{\end{equation}}
\newcommand\ea{\end{eqnarray}}

\newcommand\bw{\begin{widetext}}
\newcommand\ew{\end{widetext}}

\newcommand\nn{\nonumber}
\newcommand\mrm{\mathrm}

\newcommand\zdrift{\Delta_t z}
\newcommand\mcz{\mathcal{M}_z}

\newcommand\psiaccel{\Psi_{\mrm{acc}}(f)}

\begin{document}
\title{Probing the Inhomogeneous Universe with Gravitational Wave Cosmology}

\author{Kent Yagi}



\address{Department of Physics, Montana State University,
   Bozeman, MT, 59717}

\ead{kyagi@physics.montana.edu}

\author{Atsushi Nishizawa and Chul-Moon Yoo}

\address{Yukawa Institute for Theoretical Physics,
  Kyoto University,
  Kyoto 606--8502, Japan}

\begin{abstract}

If we assume that we live in 
the center of a spherical inhomogeneous universe, we can explain the apparent accelerating expansion of 
the universe without introducing the unknown dark energy or modifying gravitational theory.
Direct measurement of the cosmic acceleration can be a powerful tool in distinguishing $\Lambda$CDM and 
the inhomogeneous models.
If $\Lambda$CDM is the correct model, we have shown that 
DECIGO/BBO has sufficient ability to detect 
the positive redshift drift of the source
by observing gravitational waves from neutron star binaries for 5-10 years.
This enables us to rule out any Lema\^itre-Tolman-Bondi (LTB) void model 
with monotonically increasing density profile.
Furthermore, by detecting the positive redshift drift at $z\sim 0$, we can even rule out generic LTB models
unless we allow unrealistically steep density gradient at $z\sim 0$. 
We also show that the measurement accuracy is slightly improved when we consider the joint search of DECIGO/BBO and the third generation Einstein Telescope. 
This test can be performed with GW observations alone without any reference to electromagnetic observations.
\end{abstract}

\section{Introduction}

When we assume that our universe is homogeneous and isotropic, 
current cosmological observations (e.g. type Ia supernovae (SNe)~\cite{riess}) 
indicate that the cosmic expansion is accelerating. 
Once we allow 
the possibility that our universe has cosmological 
scale spherical inhomogeneity with the observer at the center, 
the observations can be explained 
without introducing the unknown dark energy or alternative theories of gravity.
In such models,  
the Copernican Principle is apparently violated and 
the cosmic expansion is not necessarily accelerating.
Therefore, future direct detection of the cosmic acceleration 
is very useful in distinguishing them. 
%
The direct detection of the acceleration of the universe 
provides not only a key to solve the dark energy problem, 
but also a critical test of the Copernican Principle. 

The simplest example of the inhomogeneous model is 
the Lema\^itre-Tolman-Bondi (LTB) spacetime 
which is a spherically symmetric 
dust
solution of the Einstein Equations.
The metric is given as
\be
ds^2 = -dt^2 + \frac{\partial_r R(t,r)^2}{1-k(r) r^2}dr^2 + R^2(t,r) d\Omega^2,
\ee
where $R(t,r)$ and $k(r)$ are arbitrary functions.
These functions are related to the density of the dust $\rho(r)$ via Einstein Equations.
If we live at the center of the LTB spacetime with a 
Gpc-scale void, 
the apparent cosmic acceleration
can be explained. 
The LTB model 
has been partially tested with the cosmological observations like, 
the cosmic microwave background, baryon acoustic oscillations, the kinetic Sunyaev-Zeldovich effect etc.,~\cite{alnes,yoo:CMB,yoo:kSZ,garcia-bellido:cGBH,garcia-bellido:kSZ,garcia-bellido:cosmicshear,bolejko,caldwell,clarksonregis,zhangstebbins,zibin,moss,nadathur}, but
it has not been completely ruled out yet. 
For example, Ref.~\cite{moss} claimed that LTB void models are in conflict with current observations with nearly scale invariant primordial spectrum, but Nadathur and Sarkar~\cite{nadathur} claim that that observational results can be explained by assuming different primordial spectrum ansatz.
%
Therefore, the observations which are unaffected by 
the primordial information are very crucial to test a wide class of LTB models. 
One of such observations is 
the redshift-distance relation of type-Ia SNe. 
%
However, it has been shown that one can construct the 
LTB void model 
that exactly reproduces the redshift-distance relation in $\Lambda$CDM~\cite{yoo:inverse}. 
Therefore, we need other observations that do not depend on the primordial information.

Redshift-drift measurement is the one that meets our 
demands~\cite{uzan:zdrift} (see also Refs.~\cite{quartin,yoo:inverse,yoo:zdrift}).
Redshift drift is the time evolution of the redshift due to 
the cosmic acceleration, hence its detection
means
the direct measurement of the 
acceleration of the cosmic expansion. 

In the Friedmann-Lema\^itre-Robertson-Walker (FLRW) spacetime, 
the redshift drift is given as
\be 
\zdrift = H_0 \Delta t_o \left( 1+z-\frac{H(z)}{H_0} \right), 
\ee
where $\Delta t_o$ denotes the observation period, 
and $H_0$ and $H(z)$ are the Hubble parameter at present 
and at redshift $z$~\cite{loeb}, respectively. 
%
%
In the $\Lambda$CDM universe, 
$\zdrift$ is \textit{positive} in the range $z=0-2$~\cite{quartin}.
On the other hand, $\zdrift$ in LTB spacetime obeys the following differential equation~\cite{yoo:zdrift}; 
\be
\frac{d}{dz}\left( \frac{\Delta_t z}{1+z} \right)= \frac{1}{(1+z)^2} \frac{\partial_t^2 \partial_r R}{\partial_t \partial_r R} \Delta t_o,
\ee
where $R(t,r)$ is an arbitrary function.
Recently, Yoo \textit{et al.}~\cite{yoo:zdrift} have shown that when the matter density is monotonically increasing, the right hand side of the above equation is negative.
By combining this with $\zdrift|_{z=0}=0$, we can show that $\zdrift$ must be negative at any $z$ in this model.
Furthermore, they have shown that for any LTB density profile, $d\zdrift/dz  < 0$
%
%
for $z \ll 1$ unless we allow unrealistically steep density profile at $z \sim 0$.
Therefore, it is crucial to measure the sign of $\zdrift$ 
at $z<2$ in order to distinguish 
the $\Lambda$CDM and LTB models. 

The order of magnitude of $\zdrift$ is roughly given as the cosmic age 
divided by the observation time, hence 
$\zdrift \sim 10^{-10} $ for 1yr observation.
It is this tiny value that makes it difficult to measure $\zdrift$ with current technology. 
Recently, 
Quartin and Amendola~\cite{quartin} have shown that 
by measuring the shift 
of the Lyman $\alpha$ forest of 
quasar spectrum at $z=2-5$ with the proposed E-ELT instrument 
CODEX~\cite{liske} for 10 yrs, it will be possible to distinguish 
$\Lambda$CDM and typical LTB void models. 
%
%
%
%
However, 
CODEX would not be able to measure $\zdrift$ at low $z$ 
since Ly$\alpha$ forest can be measured from ground only 
at $z \geq 1.7$~\cite{liske}.
Hence, they can only test typical LTB models but not generic ones.

In this paper, we estimate how accurately we can measure 
the redshift drift with future gravitational wave (GW) interferometers.
It seems that DECIGO~\cite{setoDECIGO,kawamura2011} and BBO~\cite{bbo} 
are the only proposed detectors that can measure $\zdrift$ at $z \leq 2$. 
We consider neutron star (NS) binaries as GW sources, which are often called as the standard sirens and can be unique tools to probe the cosmic expansion~\cite{schutz,nishizawa,DECIGOcosmology}.
When the expansion is accelerating, 
we may find an additional phase shift in gravitational waveforms
~\cite{setoDECIGO}. 
We assume that $\Lambda$CDM is the correct model 
and estimate whether we can tell the positivity 
of the redshift drift at low $z$ with GW observations.

Throughout this paper, we use the unit $G=c=1$.

\section{Correction in gravitational waveform due to the redshift drift}

Let us first derive the correction in the gravitational waveform phase due to the accelerating expansion of the universe.
We here consider a binary consisting of two bodies 
with masses $m_1$ and $m_2$.
We define the time to coalescence measured in the observer frame as $\Delta t \equiv t_c - t$ with $t_c$ representing the coalescence time.
This $\Delta t$ includes the effect of cosmic acceleration.
On the other hand, we denote the time to coalescence measured in the source frame as $\Delta t_e$ and define $\Delta T \equiv (1+z_c) \Delta t_e$ where $z_c$ is the source redshift at coalescence.
Then, the relation between $\Delta t$ and $\Delta T$ is~\cite{setoDECIGO,takahashinakamura} 
\be
\Delta t = \Delta T + X(z_c) \Delta T^2\,,
\ee
where $X(z)$ is the acceleration parameter defined as 
\be
X(z) \equiv \frac{H_0}{2} \left(1-\frac{H(z)}{(1+z) H_0} \right).
\ee
Notice that $X(z)$ is related to the redshift drift $\zdrift$ as 
\be
\zdrift= 2(1+z)\Delta t_o X(z).
\label{zdriftX}
\ee
By using the stationary phase approximation~\cite{cutlerflanagan}, the waveform in the Fourier domain can be expressed as
\be
\tilde{h}(f) = e^{i\psiaccel} \tilde{h}(f) \big|_{\mathrm{no \ accel}}\,,
\ee
where 
\be
\psiaccel \equiv -2\pi f X(z_c) \Delta T(f)^2
\ee
and $\tilde{h}(f) \big|_{\mathrm{no \ accel}}$ corresponds to the gravitational waveform in the Fourier domain without cosmic acceleration explained in the Appendix. 
The leading $\Delta T(f) $ is given as~\cite{cutlerflanagan} 
\be
\Delta T(f)  = 5(8\pi \mcz f)^{-8/3} \mcz, 
\ee
where  $\mathcal{M}_z\equiv M (1+z_c) \eta^{3/5}$ denotes the redshifted chirp mass with $M \equiv (m_1+m_2) $ and  $\eta \equiv m_1m_2/M^2$ representing the total mass and the symmetric mass ratio, respectively.
With this at hand, $\psiaccel $ is given as~\cite{setoDECIGO,takahashinakamura} 
\be
\psiaccel \equiv -\Psi_N (f) \frac{25}{768} X(z_c) \mcz x^{-4}\,,
\ee
where $x\equiv (\pi \mcz f)^{2/3}$.
A term proportional to $\Psi_N(f) x^n$ represents the $n$-th post-Newtonian (PN) order relative to the leading $\Psi_{N}(f)\equiv \frac{3}{128}(\pi \mathcal{M}_z f)^{-5/3}$, hence this is ``-4PN'' correction.
We used the restricted-2PN waveform including spin-orbit coupling at 1.5PN order
, where ``restricted'' means that we only take the leading Newtonian quadrupole contribution for the amplitude and neglect the ones from higher harmonics.
%
%
%
%
%
%


\begin{figure}[t]
  \centerline{\includegraphics[scale=1.6,clip]{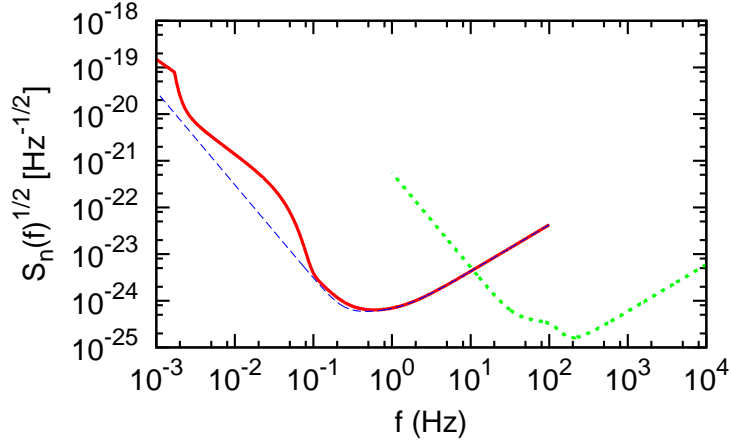} }
 \caption{\label{noise}  
The root noise spectral density of DECIGO/BBO with (solid) and without (dashed) WD confusion noises, and the one of ET (dotted).}
\end{figure}

\section{Numerical Setups}
The measurement accuracy on $\zdrift$ has been estimated mainly using ultimate DECIGO (which is three orders of magnitude more sensitive than DECIGO)~\cite{setoDECIGO,takahashinakamura}. 
Especially, Ref.~\cite{takahashinakamura} used this measurement accuracy on the acceleration parameter $X(z)$ to estimate the ones on cosmological parameters.
We improve their analyses in the following way and apply the result to probe the inhomogeneity of the universe:
(I) We use DECIGO/BBO (and not ultimate DECIGO), including the confusion noises from white dwarf (WD) binaries.
(II) Rather than sky-averaged analysis, we perform Monte Carlo simulations by randomly distributing the directions and orientations of sources. 
(III) We include the spin-orbit coupling into binary parameters.
(IV) We use the merger rate that reflects the star formation history.
%
%

We take the binary parameters as
\be
\theta^i = \left( \ln \mcz, \ln \eta, \beta, t_c, \phi_c, D_L, \bar{\theta}_{\mrm{S}}, \bar{\phi}_{\mrm{S}}, \bar{\theta}_{\mrm{L}}, \bar{\phi}_{\mrm{L}}, X_H \right).
\ee
Here, $\beta$, $\phi_c$ and $D_L$ represent the spin-orbit coupling parameter, the coalescence phase and the luminosity distance, respectively.
$(\bar{\theta}_{\mrm{S}}, \bar{\phi}_{\mrm{S}})$ are the direction of the source in barycentric frame which is tied to the ecliptic and centered in the solar system barycenter (see Fig. 1 of Ref.~\cite{yagiLISA}), and $(\bar{\theta}_{\mrm{L}}, \bar{\phi}_{\mrm{L}})$ are the orientation of the source orbital axis in the same frame.
We have introduced a new parameter $X_H$ which is defined as $X_H \equiv X(z)/H_0$. 
$X_H$ is related to $\zdrift$ through Eq.~(\ref{zdriftX}).

We estimate how accurately we can measure binary parameters $\theta^i$ using Fisher analysis.
Assuming that the detector noise is stationary and Gaussian, the measurement accuracy is given as $ \Delta\theta^i  \equiv \left( \tilde{\Gamma}^{-1} \right)^{1/2}_{ii}$~\cite{cutlerflanagan},
%
%
where, $\tilde{\Gamma}_{ij}$ is defined as
\be
\exp\left[ -\frac{1}{2}\tilde{\Gamma}_{ij}\delta\theta^i\delta\theta^j \right] \equiv p^{(0)}(\bm{\theta})\exp\left[ -\frac{1}{2}\Gamma_{ij}\delta\theta^i\delta\theta^j \right].
\label{fisher}
\ee
Here, $\delta \theta^i \equiv \theta^i-\theta^i_\mrm{true}$ with $\theta^i_\mrm{true}$ representing the true values of binary parameters and $p^{(0)}(\bm{\theta})$ is the prior information. 
$\Gamma_{ij}$ is the Fisher matrix which is defined as
\be
\Gamma_{ij} \equiv 4 \mathrm{Re}\int ^{f_{\mrm{fin}}}_{f_{\mrm{in}}}df \, \frac{\tilde{h}^{*}(f)\tilde{h}(f)}{S_n(f)}, 
\label{scalar-prod}
\ee
where $S_n(f)$ denotes the noise spectrum of the detector.
In this paper, we use the instrumental noise spectrum of BBO when we perform Fisher analyses and assume that DECIGO also has the same sensitivity as BBO.
The noise curves are shown in Fig.~\ref{noise}.
The actual expression of the total noise spectrum is given in Eq.~(36) of Ref.~\cite{yagi:brane}.

$f_{\mrm{in}}$ and $f_\mrm{fin}$ in Eq.~(\ref{scalar-prod}) are given as~\cite{yagi:brane} 
\be
f_{\mrm{in}}=(256/5)^{-3/8} \pi^{-1} \mcz^{-5/8} \Delta t_o^{-3/8}, \qquad f_{\mrm{fin}}=100\mrm{Hz}, 
\ee
respectively. 
$f_{\mrm{in}}$ represents the frequency at $\Delta t_{o}$ before coalescence and $f_\mrm{fin}$ denotes the higher cutoff frequency of DECIGO/BBO.
For fiducial values, we set $m_1=m_2= 1.4M_{\odot}$ and take $t_c=\phi_c=\beta=0$.
Since the dimensionless spin parameter is less than 1, we adopt the prior as $|\beta| < 9.4$~\cite{bertibuonanno}.
We assume that a flat $\Lambda$CDM is the correct model.
In the following computations, we set the fiducial values as $H_0=70$km/s/Mpc and the cosmological parameters as $\Omega_m=0.3$ and $\Omega_{\Lambda}=0.7$.

The number of binaries $\Delta N(z)$  that exists in each bin with size $\delta z = 0.1$ is estimated as~\cite{cutlerharms} 
\be
 \Delta N(z) =  4\pi \left[ a_0 r(z) \right]^2 \dot{n}(z) (d\tau /dz) \delta z \Delta t_o,
\ee
 where 
\be
a_0r(z) =\int ^z_0 dz'/H(z'), \qquad {d\tau}/{dz} = \{ (1+z)H(z) \}^{-1}
\ee
 with the Hubble parameter at redshift $z$ given as
\be
H(z) \equiv H_0 \sqrt{\Omega_m (1+z)^3+\Omega_{\Lambda}}.
\ee
Here, $a_0$, $r(z)$, and $\tau$ each represents current scale factor, comoving distance and proper look back time, respectively. 
$\dot{n}(z) \equiv \dot{n}_0 R(z)$ shows the NS/NS merger rate per unit comoving volume per unit proper time, where we assume the merger rate today as $\dot{n}_0=10^{-6}$ Mpc$^{-3}$ yr$^{-1}$ ~\cite{abadie} and the merger rate evolution against $z$ as
%
\ba
R(z)=\left\{ \begin{array}{ll}
1+2z & (z\leq 1) \\
\frac{3}{4}(5-z) & (1\leq z\leq 5) \\
0 & (z\geq 5), \\
\end{array} \right.
\ea
which is based on the current observation of star formation history~\cite{schneider}.

Following Refs.~\cite{bertibuonanno,yagiLISA,yagi:brane}, we randomly generate $10^4$ sets of $(\bar{\theta}_{\mrm{S}},\bar{\phi}_{\mrm{S}},\bar{\theta}_{\mrm{L}},\bar{\phi}_{\mrm{L}})$ for each fiducial redshift $z_k \equiv 0.1k+0.05 \ (k=0,1,\cdots)$.
Then, we calculate $\left( \tilde{\Gamma}^{-1} \right)^{1/2}_{ii}$ for each set and take the average to yield $\left[\left( \tilde{\Gamma}^{-1} \right)^{1/2}_{ii}\right]_{\mrm{ave}}$. 
The measurement accuracy at each $z_k$ is estimated as
%
\be
\Delta\theta^i  = N_{\mrm{int}}^{-1/2} \Delta N(z_k)^{-1/2}  \left[\left( \tilde{\Gamma}^{-1} \right)^{1/2}_{ii}\right]_{\mrm{ave}},
\ee
where $N_{\mrm{int}}=8$ shows the number of effective interferometers.  
DECIGO/BBO consists of four clusters of triangular detectors (see e.g. Fig. 2 of Ref.~\cite{yagi:brane} for the proposed configurations of DECIGO/BBO) and for simplicity, we assume that all the clusters are placed on the same site.
We also use the sky-averaged analysis to see how the new effects that we considered in this paper affect our results.

\section{Results}

\begin{figure}[t]
  \centerline{\includegraphics[scale=1.6,clip]{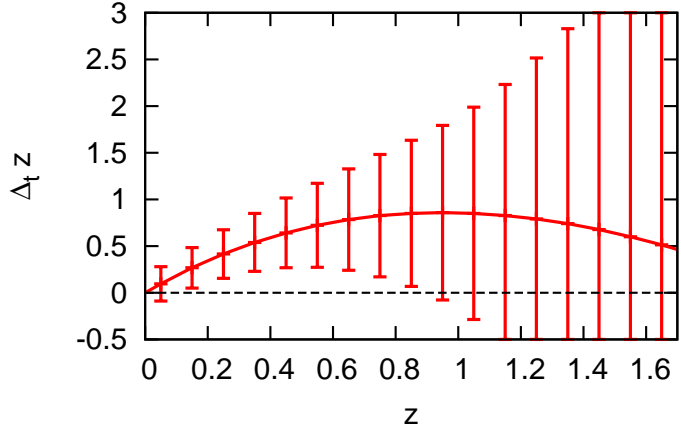} }
 \caption{\label{drift-5}  
The measurement accuracies of $\zdrift$ using DECIGO/BBO for 5 yr observations.}
\end{figure}

\begin{figure}[h]
  \centerline{\includegraphics[scale=1.6,clip]{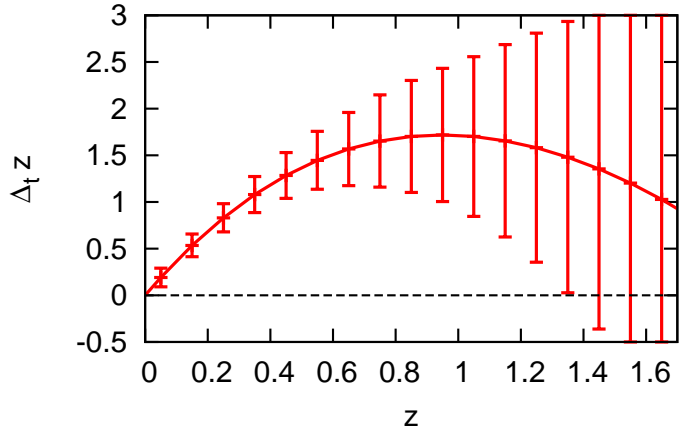} }
 \caption{\label{drift-10}  
The same as Fig.~\ref{drift-5} but the observation period changed to 10 years.}
\end{figure}

In Fig.~\ref{drift-5}, we show the measurement accuracies of $\zdrift$ using DECIGO/BBO for 5 yr observations. 
We see that the positivity of $\zdrift$ can be marginally detected at redshift around $z\sim 0.5$.
Figure~\ref{drift-10} is same as Fig.~\ref{drift-5} but for 10 yr observations. 
If this observational period is realized, $\zdrift > 0$ can be detected with 3-$\sigma$ confidence level.

\begin{figure}[t]
  \centerline{\includegraphics[scale=1.6,clip]{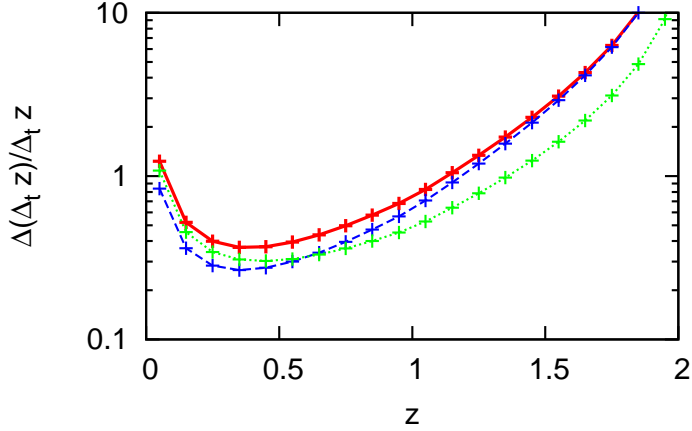} }
 \caption{\label{drift-skyave}  
$\Delta (\zdrift) / \zdrift$ using sky-averaged analysis for 5 yr observations, taking both $\beta$ and WD confusion noise into account (solid), without $\beta$ (dashed) or without WD confusion noise (dotted). }
\end{figure}

Next, we show the effects of including $\beta$ and WD confusion noises.
Figure~\ref{drift-skyave} shows $\Delta (\zdrift) / \zdrift$ for the sky-averaged analyses with 5 yr observations where the sky-averaged waveform is given in Eq.~(\ref{sky-averaged-waveform})~\footnote{We have assumed multiple detectors, but in principle, sky-averaged analysis can be performed with a single detector.}
We see that the spin-orbit coupling $\beta$ affects the result on lower $z$ side.
This is because the redshifted mass $M_z$ is lower for lower $z$ source, which leads to smaller orbital velocity $v$ at a given frequency $f$ and larger degeneracies between $\beta$ and other binary parameters.
On the other hand, WD confusion noise affect higher $z$ side.
This is because higher $z$ source leads to GW signal with lower frequency where the effect of the confusion noise is larger. 
   
\begin{figure}[t]
  \centerline{\includegraphics[scale=1.6,clip]{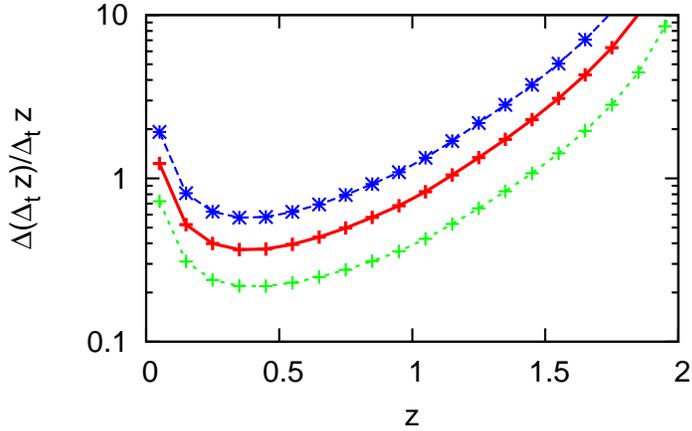} }
 \caption{\label{drift-skyave-mc}  
Solid curve is same as the one in Fig.~\ref{drift-skyave} while dotted curve does not include $\beta$ nor WD confusion noise. Dashed one shows the result with the Monte Carlo simulation. }
\end{figure}

Similarly, let us compare the results using sky-averaged analysis and Monte Carlo simulation.
Figure~\ref{drift-skyave-mc} shows the former in solid and the latter in dashed lines.
Also, sky-averaged analysis without taking $\beta$ nor WD confusion is shown in dotted line.
By comparing the dashed curve with this dotted one, we see that taking $\beta$, WD confusion noise, directions and orientations of sources into account deteriorate the determination accuracies of $\zdrift$ by a factor of a few.
Since 5 yr observation can only marginally detect the positivity of $\zdrift$, this difference is crucial in detecting $\zdrift > 0$.

\begin{figure}[t]
  \centerline{\includegraphics[scale=1.6,clip]{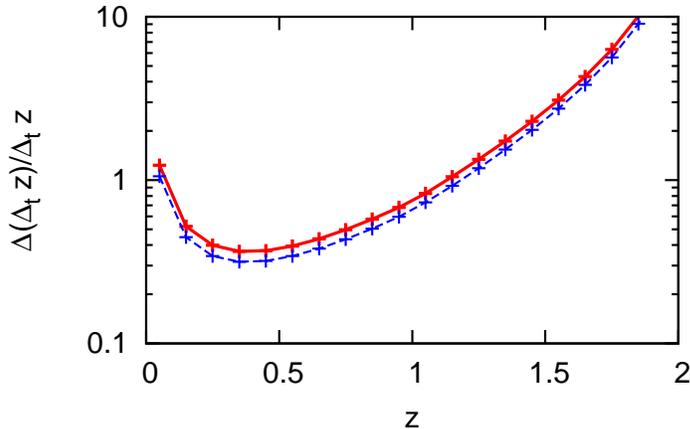} }
 \caption{\label{joint}  
$\Delta (\zdrift) / \zdrift$ using sky-averaged analysis for 5 yr observations using DECIGO/BBO only (solid), and with joint searches of DECIGO/BBO and ET (dashed). }
\end{figure}

\section{Conclusions and Discussions}

In this paper, we have estimated how accurately 
we can directly measure the cosmic acceleration using future space-borne GW interferometers such as DECIGO and BBO.
If we assume that the $\Lambda$CDM model is correct,
we have shown that we will be able to measure the positivity of the redshift drift with 5-10 yr observations, 
which enables us to rule out \textit{any} LTB void model 
with monotonically increasing density profile.
Furthermore, since 10 yr observation allows us to detect the positive redshift drift at $z\sim 0.05$, it seems that we can even rule out generic LTB void models unless we allow unrealistic density profile at $z \sim 0$. 
In order to measure the positivity of $\zdrift$, it is more useful to consider the accumulated $\zdrift$~\cite{yagi-zdrift}.

In this paper, we have assumed that the correct model is $\Lambda$CDM but it can be other dark energy model.
However, it has been shown that a variety of dark energy models predict the similar values of $\zdrift$ 
to $\Lambda$CDM one (see Quartin and Amendola~\cite{quartin} and references therein).
Therefore, we emphasize that our results are not restricted to $\Lambda$CDM only.

Unfortunately, it seems very difficult to measure $\zdrift$ 
with the ground-based detectors, even with the third-generation ones like ET~\cite{et}. 
(The noise curve is shown as dotted curve in Fig.~\ref{noise}.)
The main reasons for this is because it has less number of GW cycles, the event rate is smaller, and the frequency is higher which leads to smaller contribution of "-4PN" effect.
However, ET can help improving the measurement accuracy of $\zdrift$ when combined with DECIGO/BBO.
Figure~\ref{joint} shows the results comparing the measurement accuracy of $\zdrift$ with DECIGO/BBO observation only (solid), and the one obtained by the joint search of DECIGO/BBO \textit{and} ET (dashed.)
We see that the errors are slightly reduced for the latter case. 

There can be peculiar acceleration of each binary which acts as
an additional ``noise'' 
when measuring the cosmic acceleration. 
However, Amendola \etal\cite{amendola:peculiar} have estimated the peculiar 
accelerations for typical clusters and galaxies, 
and found that they are almost the same 
magnitude as the cosmological acceleration. 
While the peculiar acceleration is different for each binary source, the cosmic acceleration is a universal effect.
As such, 
we can safely neglect the effect of peculiar accelerations as a noise source since the one from the detector overwhelms it.
(See also Uzan \etal\cite{uzan:peculiar} for the discussion about the peculiar acceleration.)

In this paper, we have assumed that the 
binary orbits
are circular. 
Once we include the eccentricity into parameters, 
this may have large degeneracy with $\zdrift$. 
We need to estimate how the measurement accuracy of $\zdrift$ is reduced for the eccentric binaries in future.
Also, we note here that we have used 
the LTB metric as a simple effective model. 
To be more realistic, we need to use more sophisticated models such as Swiss-Cheese model~\cite{swisscheese} (see a recent discussion by C\'el\'erier~\cite{celerier}). 
We leave these issues for future work.

\ack

We thank Takahiro Tanaka, Naoki Seto, Takashi Nakamura and Daniel Holz for having discussions and giving us valuable comments.
KY~is supported by the Japan Society for the Promotion of Science (JSPS) grant No.~$22\cdot 900$.
AN~and CY is supported by a Grant-in-Aid through JSPS.
This work is also supported in part by the Grant-in-Aid for the Global COE Program ``The Next Generation of Physics, Spun from Universality and Emergence'' from the Ministry of Education, Culture, Sports, Science and Technology (MEXT) of Japan.


\appendix

\section{Binary gravitational waveform without the effect of cosmic acceleration}
\label{waveform}

The restricted binary gravitational waveform in Fourier space without cosmic acceleration is given as~\cite{cutler1998,bertibuonanno}
\be
\tilde{h}(f)|_\mrm{no \ accel}=\frac{\sqrt{3}}{2}\mathcal{A}f^{-7/6}e^{i\Psi (f)} \left[ \frac{5}{4}A_{\mathrm{pol},\alpha}(t(f)) \right] e^{-i \left( \varphi_{\mathrm{pol},\alpha}+\varphi_D \right)}, 
\label{waveform}
\ee
%
where the amplitude $\mathcal{A}$ is given as $\mathcal{A}=\frac{1}{\sqrt{30}\pi^{2/3}}\frac{\mathcal{M}_z^{5/6}}{D_L}$ and
the phase up to second Post-Newtonian (PN) order can be expressed as~\cite{setoDECIGO,bertibuonanno}
%
\ba
\Psi (f) &= & 2\pi ft_c-\phi_c -\frac{\pi}{4}+ \Psi_N(f)   \times  \bigg[1 + \left( \frac{3715}{756}+\frac{55}{9}\eta \right)x  -4(4\pi-\beta)x^{3/2} \nn \\
        & &   + \left( \frac{15293365}{508032}+\frac{27145}{504}\eta+\frac{3085}{72}\eta^2 \right) x^2  \bigg]. \nn \\
\label{Psi-noangle}
\ea
%
Here, $\phi_c$ is the coalescence phase and
$A_{\mathrm{pol},\alpha}$,  $\varphi_{\mathrm{pol},\alpha}$ and $\varphi_D$ are given as~\cite{bertibuonanno,yagiLISA}
\ba
A_{\mathrm{pol},\alpha}(t)&=&\sqrt{(1+(\hat{\bm{L}}\cdot\hat{\bm{N}})^2)^2F_{\alpha}^{+}(t)^2+4(\hat{\bm{L}}\cdot\hat{\bm{N}})^2F_{\alpha}^{\times}(t)^2}, \label{Apol} \\
 \cos(\varphi_{\mathrm{pol},\alpha}(t))&=&\frac{(1+(\hat{\bm{L}}\cdot\hat{\bm{N}})^2)F^{+}_{\alpha}(t)}{A_{\mathrm{pol},\alpha}(t)}, \\
 \sin(\varphi_{\mathrm{pol},\alpha}(t))&=&\frac{2(\hat{\bm{L}}\cdot\hat{\bm{N}}) F^{\times}_{\alpha}(t)}{A_{\mathrm{pol},\alpha}(t)}, \label{phipol} \\
\varphi_{D}(t)&=&2\pi f(t) R \sin \bar{\theta}_{\mathrm{S}} \cos[2\pi t/T-\bar{\phi}_{\mathrm{S}}], \label{doppler-phase}
\ea
where $\hat{\bm{L}}$ is the unit vector parallel to the orbital angular momentum, $\hat{\bm{N}}$ is the unit vector pointing towards the center of mass of the binary, $T$=1yr and $R$=1AU.
$F_{\alpha}^{+}$ and $F_{\alpha}^{\times}$ are the beam pattern functions defined as 
\ba
F_{\mathrm{I}}^{+}(\theta_{\mathrm{S}},\phi_{\mathrm{S}},\psi_{\mathrm{S}}) 
                &=&\frac{1}{2}(1+\cos^2 \theta_{\mathrm{S}}) \cos(2\phi_{\mathrm{S}}) \cos (2\psi_{\mathrm{S}})
                  -\cos(\theta_{\mathrm{S}}) \sin(2\phi_{\mathrm{S}}) \sin(2\psi_{\mathrm{S}}),  \\
F_{\mathrm{I}}^{\times}(\theta_{\mathrm{S}},\phi_{\mathrm{S}},\psi_{\mathrm{S}})
                &=&\frac{1}{2}(1+\cos^2 \theta_{\mathrm{S}}) \cos(2\phi_{\mathrm{S}}) \sin (2\psi_{\mathrm{S}})
                  +\cos(\theta_{\mathrm{S}}) \sin(2\phi_{\mathrm{S}}) \cos(2\psi_{\mathrm{S}}), \\
F_{\mathrm{II}}^{+}(\theta_{\mathrm{S}},\phi_{\mathrm{S}},\psi_{\mathrm{S}})&=&F_{\mathrm{I}}^{+}(\theta_{\mathrm{S}},\phi_{\mathrm{S}}-\pi/4,\psi_{\mathrm{S}}), \\
F_{\mathrm{II}}^{\times}(\theta_{\mathrm{S}},\phi_{\mathrm{S}},\psi_{\mathrm{S}})&=&F_{\mathrm{I}}^{\times}(\theta_{\mathrm{S}},\phi_{\mathrm{S}}-\pi/4,\psi_{\mathrm{S}}),
\ea
where the direction $(\theta_\mrm{S},\phi_\mrm{S})$ is measured in the detector's frame and $\psi_\mrm{S}$ is the polarization angle.
These angles and $\hat{\bm{L}}\cdot \hat{\bm{N}}$ are related to the angles $(\bar{\theta}_\mrm{S},\bar{\phi}_\mrm{S},\bar{\theta}_\mrm{L},\bar{\phi}_\mrm{L})$ measured from the center of mass of the solar system as explained in Appendix A of Ref.~\cite{yagiLISA}.
%
We have neglected the spin-spin coupling at 2PN order since it has been shown that this effect is negligible for NS/NS binaries~\cite{cutlerharms}.
%
%
%
$t(f)$ in Eq.~(\ref{waveform}) up to 2PN order is given as
%
\ba
t(f) &=& t_c-t_N(f) \bigg[1 +\frac{4}{3}\left( \frac{743}{336}+\frac{11}{4}\eta \right) x  -\frac{8}{5}(4\pi-\beta) x^{3/2} \nn \\
  & &     +2\left( \frac{3058673}{1016064}+\frac{5429}{1008}\eta+\frac{617}{144}\eta^2 \right) x^2 \bigg], \label{tf}
\ea
%
with $t_N(f) \equiv (5/256) \mcz (\pi \mcz f)^{-8/3}$~\footnote{When we consider the effect of cosmic acceleration, there appears a leading ``-4PN'' correction in $t(f)$, but we have checked that it does not affect our results.}.
When we take the average over the direction and orientation of sources, Eq.~(\ref{waveform}) just reduces to
\be
\tilde{h}(f)|_\mrm{no \ accel}^\mrm{(sky-averaged)}=\frac{\sqrt{3}}{2}\mathcal{A}f^{-7/6}e^{i\Psi (f)} . 
\label{sky-averaged-waveform}
\ee

\section*{References}

\bibliography{ref}

\end{document}